\documentclass{article}

\usepackage{arxiv}

\usepackage[utf8]{inputenc} 
\usepackage[T1]{fontenc}    
\usepackage{url}            
\usepackage{booktabs}       
\usepackage{amsfonts}       
\usepackage{nicefrac}       
\usepackage{microtype}      
\usepackage{lipsum}         
\usepackage{graphicx}
\usepackage{natbib}
\usepackage{doi}

\usepackage[section]{placeins}
\usepackage{amsmath}
\usepackage{amssymb}
\usepackage{grfext}
\usepackage[linesnumbered,ruled,vlined,onelanguage]{algorithm2e}
\usepackage{xcolor}
\usepackage{algpseudocode}
\usepackage{float}
\usepackage{parskip}
\usepackage{enumitem}
\setlist[itemize]{leftmargin=*}
\usepackage{csvsimple}
\usepackage{caption}
\usepackage{subcaption}

\usepackage{lineno}
\title{Predictive Data Calibration for Linear Correlation Significance Testing}

\author{ 
    \href{https://orcid.org/0000-0000-0000-0000}{\includegraphics[scale=0.06]{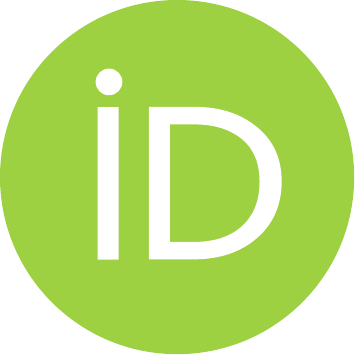}\hspace{1mm}Kaustubh R.~Patil\textsuperscript{1,2}}\\
	\texttt{k.patil@fz-juelich.de}\\
	\And
	\href{https://orcid.org/0000-0000-0000-0000}{\includegraphics[scale=0.06]{orcid.pdf}\hspace{1mm}Simon B.~Eickhoff\textsuperscript{1,2}}\\
	\texttt{s.eickhoff@fz-juelich.de}\\
	\And
	\href{https://orcid.org/0000-0000-0000-0000}{\includegraphics[scale=0.06]{orcid.pdf}\hspace{1mm}Robert Langner\textsuperscript{1,2}}\\
	\texttt{r.langner@fz-juelich.de}\\
	\\
	\textsuperscript{1}Institute of Neuroscience and Medicine (INM-7: Brain and Behaviour)\\
	Forschungszentrum Jülich\\
	52425 Jülich, Germany\\
	\\
	\textsuperscript{2}Institute of Systems Neuroscience\\
	Medical Faculty, Heinrich Heine University\\
	40225 Düsseldorf, Germany\\	
}

\renewcommand{\shorttitle}{\textit{Patil et al.} Predictive Data Calibration}

\begin{document}
\maketitle

\begin{abstract}
Inferring linear relationships lies at the heart of many empirical investigations.
A measure of linear dependence should correctly evaluate the strength of the relationship as well as qualify whether it is meaningful for the population. 
Pearson's correlation coefficient (PCC), the \textit{de-facto} measure for bivariate relationships, is known to lack in both regards. 
The estimated strength $r$ maybe wrong due to limited sample size, and nonnormality of data.
In the context of statistical significance testing, erroneous interpretation of a $p$-value as posterior probability leads to Type I errors---a general issue with significance testing that extends to PCC.
Such errors are exacerbated when testing multiple hypotheses simultaneously. 
To tackle these issues, we propose a machine-learning-based predictive data calibration method which essentially conditions the data samples on the expected linear relationship. 
Calculating PCC using calibrated data yields a calibrated $p$-value that can be interpreted as posterior probability together with a calibrated $r$ estimate, a desired outcome not provided by other methods.  
Furthermore, the ensuing independent interpretation of each test might eliminate the need for multiple testing correction. 
We provide empirical evidence favouring the proposed method using several simulations and application to real-world data.
\end{abstract}


\section{Introduction}

Inferring linear relation between two observed variables is a ubiquitous task in empirical sciences. 
The century old Pearson’s correlation coefficient (PCC) \citep{pearson_mathematical_1896} has become the \textit{de-facto} measure to estimate bivariate linear relationships.
A PCC test provides an estimate of the strength and direction of the relationship $r$ together with a $p$-value.
In the framework of Null Hypothesis Significance Testing (NHST), this $p$-value quantifies the uncertainty associated with observing a linear association ($H1$) when the null hypothesis ($H0$) of no association were true. 
A $p$-value lower than a predefined point-wise significance level $\alpha$ (usually set to $0.05$) is taken as evidence for rejecting $H0$ and $H1$ is declared to be true.
This significance testing procedure is widely adopted due to its ease of application and domain-independent nature. 

PCC, however, is not perfect and both the estimated strength $r$ and $p$-value suffer from shortcomings. First, violation of assumptions behind PCC, such as presence of outliers and nornormality of data, leads to mischaracterization of the linear dependence \citep{pernet_robust_2013,bishara_reducing_2015,armstrong_pearson_2019}.
Second, erroneous interpretation of a $p$-value as the probability of $H0$ being true leads to inflated Type I errors and has been criticised of its interpretation and reliability \citep{goodman_dirty_2008,gao_p-values_2020}, and blamed for the replication/reproducibility crisis \citep{nuzzo_scientific_2014, halsey_fickle_2015, anderson_misinterpreting_2020}. 
In a wider context there are calls to abandon NHST altogether \citep{szucs_ioannidis_NHST_2017,haggstrom_need_2017,cumming_newstats_2014}. 
An associated issue is that of \textit{multiple testing} (also called multiplicity) which leads to a surge in the Type I error rate (increased false positives) when multiple tests are performed simultaneously. 
With large-scale data measuring thousands or even millions of variables becoming common owing to the advancements in data collection techniques and falling costs, multiple testing has become a common concern.

To alleviate the issue of incorrect estimation of the $r$ value, there have been suggestions of alternative or specialized methods that exhibit robustness \citep{pernet_robust_2013,bishara_reducing_2015,wilcox_robust_2012}. 
However, such methods are often designed for handling specific issues, for instance robustness to outliers using skipped-correlations \citep{wilcox_skipped_2015}.
Use of alternate methods, rank-based Spearman's correlation can be used which is less affected by outliers but measures monotonic dependence and not linear dependence as desired here. 
Other options include performing non-linear transformation of the data, e.g. Box-Cox or rankit \citep{bishara_reducing_2015}.
 
Regarding the (mis)use of $p$-values, their reflected use has been advised with a caution that any interpretation alternative will also be susceptible to similar fallacies \citep{leek_peng_statistics_2015,nickerson_null_2000,gibson_role_2021,lakens_practical_2021} (also see statement from the American Statistical Association \citep{wasserstein_asa_2016}). 
It has been argued that point estimates can provide valuable insights and are an essential part of science for testing a claim \citep{mayo_statistical_2022}.
A possible remedy is \textit{"making $p$ values work harder"} \citep{matthews_2021} (cf. March 2019 special issue of The American Statistician \citep{wasserstein_moving_2019}), e.g. estimating posterior probabilities by $p$ value calibration \citep{sellke_calibration_2001,shi_reconnecting_2021,bickel_null_2021,cabras_p-value_2017,gao_p-values_2020,rafi_greenland_semantic_2020}. After this calibration the type I error probability can be interpreted as as the probability that the null hypothesis is true. As $p$ value calibration is easy to compute and interpret it provides an attractive way to counter misinterpretation of $p$ values.

To deal with the multiple testing, correction to the $p$-values arising from the individual tests is commonly employed. 
Several correction methods have been proposed which rely on the number of tests and/or the distribution of $p$-values. 
These methods differ in their strictness in penalizing $p$-values reflecting the trade-off between false positives and false negatives, e.g. family-wise error rate (FWER) correction provides the strictest control while false discovery rate (FDR) control is more lenient. 
However, any $p$-value correction method is not without issues and choosing a correction method is usually left to the users. 
As the correction is often proportional to the number of tests, the power to identify an effect decreases with the number of tests. 
Defining a domain for control is not straightforward, e.g. whether to exert control over data used within an experiment or an entire discipline \citep{trafimow_earp_2017}.
Dependencies between the tests (usually not known \emph{a priori}) require special consideration as ignoring them may increase false negatives \citep{leek_general_2008,wilson_harmonic_2019}. 
Furthermore "\emph{big data}" with number of tests ranging in hundreds of thousands or even millions, even lenient correction can lose power. 
General methods like local FDR \citep{efron_empirical_2001} as well as domain-specific methods with higher power such as threshold free cluster enhancement (TFCE) \citep{smith_threshold-free_2009} for MRI images and independent hypothesis weighting (IHW) for genomics \citep{ignatiadis_data-driven_2016} have been developed. 
However, specialized methods can not be immediately applied to other fields. 
Taken together, the issues related to data assumptions, $p$-value interpretation and multiple testing make the trivial sounding task of inferring linear relationships rather challenging and new methods that can solve one or more of the issues are in demand.

If a hypothesis testing method could provide a $p$-value that can be interpreted as the probability of the null hypothesis being true and in isolation, i.e. independent of other tests, then the above mentioned issues would be resolved. 
Specifically, such a test (1) would not need multiple testing correction, and (2) it will not be susceptible to dependencies between the hypotheses. 
A posterior probability estimate can indeed serve this purpose \citep{}. 
Here we take steps in this direction and propose a method which provides an alternative way to perform large scale statistical tests without having to do multiple testing corrections. 
The proposed correction-free method rests on the rationale that if a test can be declared generally significant then the associated $p$-value can be used directly for inference and does not require correction. 
By generally we mean if the test is repeated on other samples of data the it would yield a similar result. 
This implicit generalizability makes such the test independent of other tests and renders correction unnecessary, paving the way towards high-throughput analysis by identifying as many significant effects as possible, while producing relatively few false positives. 
In this article we propose a new machine-learning-based method for \textit{predictive data calibration}.

We propose a new method where each test is treated independently by combining the notion and estimation of generalization as defined in the machine-learning field with inferential statistics. 
Here, we adopt the cross-validation procedure aimed at estimating the out-of-sample (i.e. generalization) prediction performance based on a finite sample. 
This is achieved by training a predictive model (e.g. linear regression) on a subset of the data and using the remaining subset to evaluate the prediction performance. 
The out-of-sample predictions then provide a generalized representation of the predicted variable. 
Thus, performing the statistical tests on the variables in their generalized forms provides an inference mechanism which can be interpreted as being independent of the other tests.
In the proposed data calibration method \texttt{dcal} we replace each value with its out-of-sample prediction. 
The proposed method reduces the power of the test in a data-driven way solely based on the data of a particular test as opposed to traditional multiple testing correction where the correction is typically dependent on the number of tests.
We claim that the proposed approach captures the \emph{true} relationship and hence provides protection against the multiple comparison problem. 
Indeed we empirically show this to be the case by performing numerical experiments on simulated and real-world data.

\section*{Predictive Data Calibration: Proposal and Design}
Our proposed method, which we call data calibration (\texttt{dcal}), relies on estimating \emph{generalizable information}. \texttt{dcal} works by calibrating the data to the alternative hypothesis under test (following the convention that the default hypothesis is the null of no relationship) using out-of-sample prediction \citep{hastie_elements_2009}. In effect, it regularizes the data such that the samples are penalized by its deviation--i.e. the prediction error it incurs--w.r.t. the hypothesis. To this end, \texttt{dcal} uses a learning algorithm to replace each sample by their out-of-sample (OOS) prediction, in effect creating new data calibrated to the alternative hypothesis and then applies the classical test to this new calibrated data.

Here we focus on linear correlation and propose an alternative to the classical linear correlation test. Linear correlation is calculated for two random variables $x\;(x_1,x_2,\dots,x_n \in \mathbb{R})$ and $y\;(y_1,y_2,\dots,y_n \in \mathbb{R})$ which provides a correlation coefficient and a $p$-value $r,p=cor(x,y)$. This $p$-value is then used as the basis for NHST. 

For linear correlation, linear regression (LR) fitted using the ordinary least squares (OLS) is the natural choice as a learning algorithm. Thus for each of the two variables we calculate mutual out-of-sample predictions for each value, e.g. using the leave-one-out (LOO) cross-validation procedure:

\begin{equation}
\label{eq1}
\begin{split}
\hat{y_i} = OLS^{(y \sim x)_{-i}}(x_i) \\
\hat{x_i} = OLS^{(x \sim y)_{-i}}(y_i)
\end{split}
\end{equation}

where $-i$ indicates a subset without the $i^{th}$ sample and the corresponding $OLS$ is a model induced using this subset. In words, we induce a linear regression function by leaving out a particular instance and estimate it's value by predicting it. We then calculate the linear correlation between the predicted variables which provides the \texttt{dcal} version of the linear correlation test. 

\begin{equation} \label{eq2}
r_{dcal}, p_{dcal} = cor(\hat{x}, \hat{y})
\end{equation}

\begin{figure}[ht]
\centering
\includegraphics[width=0.8\linewidth,keepaspectratio]{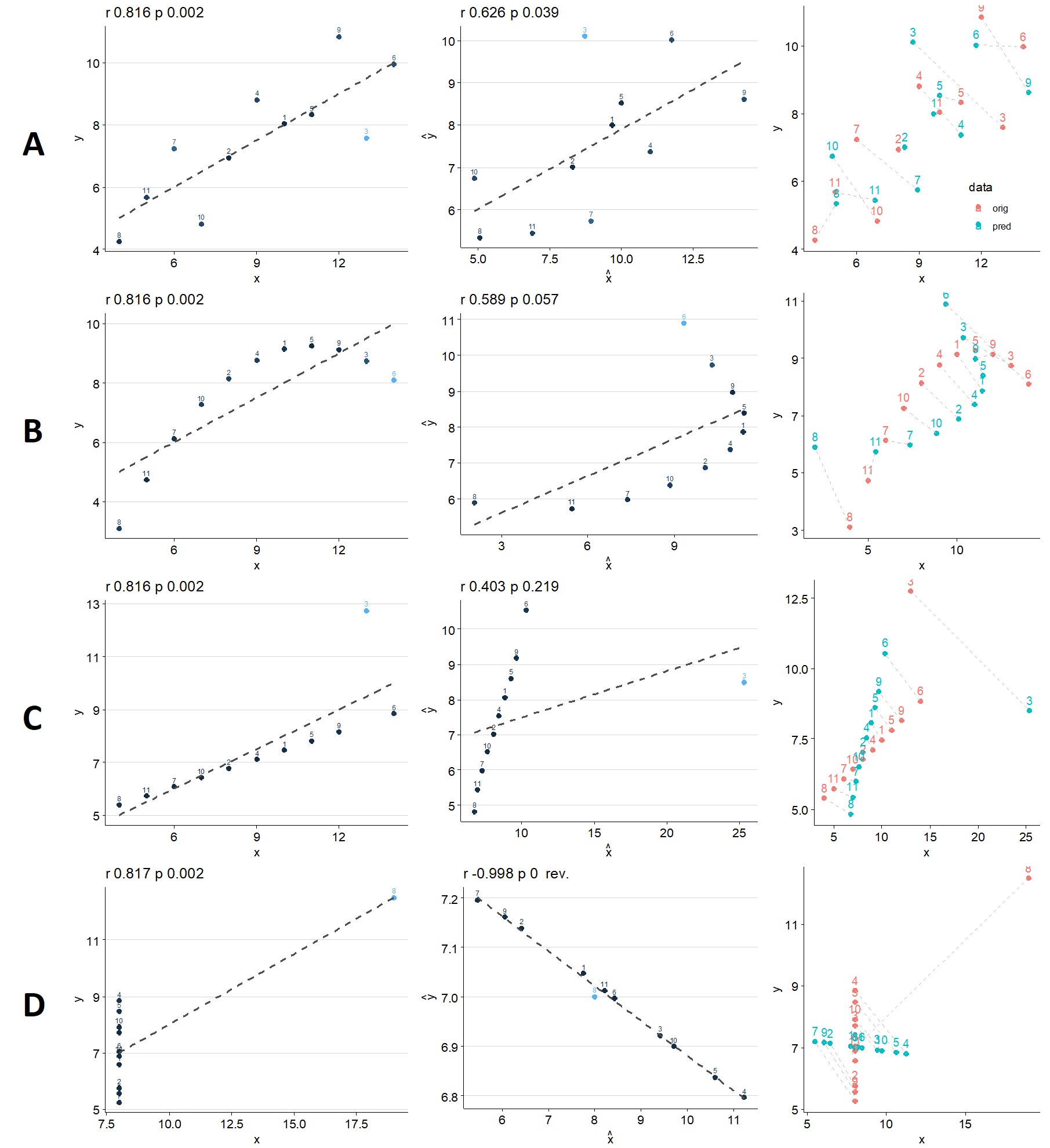}
\caption{Analysis of the Anscombe's quartet. The four rows show one dataset each (A-D) wherein the scatter plot on the left shows the linear correlation and the middle scatter plot side shows \texttt{dcal} linear correlation. The dashed lines show linear fits. The right panels show original predicted values and their relationship.}
\label{fig_anscombe}
\end{figure}

An illustrative and intuitive way to showcase the usefulness of the proposed method is using the Anscombe's quartet \citep{anscombe_graphs_1973}. 
This dataset was created to demonstrate the pitfalls when using the linear correlation and consists of four pairs of variables with identical first order statistical properties and the same linear correlation but the actual nature of relationship varies widely. 
By design, only the first variable pair is valid for linear correlation while the other three violate the assumptions of linear correlation. 
The established $p$-value calibration methods fail to detect the three invalid datasets as all datasets show a $p$-value $\sim 0.0022$ which after application of the calibration method proposed by Selke and colleagues \citep{sellke_calibration_2001} will yield a calibrated $p$-value of $\sim 0.035$.
Similarly, Bayesian correlation is also not immune to these pitfalls \citep{van_doorn_jasp_2020} while robust correlation analysis with possible manual visual checks can revel some of the issues \citep{pernet_robust_2013}. 
Overall, although simplistic, analysis of the Anscombe's quartet revealed several illuminating findings (Fig. \ref{fig_anscombe}).

\begin{quote}
\begin{itemize}
\item The \texttt{dcal} results were different for the four datasets while, as expected, they were the same for the classical correlation tests.
\item For datasets A-C, the \texttt{dcal} correlation values were lower and the $p$ values values higher than their classical counterparts.
\item The $r_{dcal}$ values decreased and the $p_{dcal}$ values increased from A to C as the violations by the data become more severe.
\item Datasets B and C did not result in a significant \texttt{dcal} test with the traditionally used threshold of $p < 0.05$ (B would be significant at a trend-level $p<0.1$).
\item For dataset D, the sign of the \texttt{dcal} correlation reversed with a higher $r_{dcal}$ and a lower $p_{dcal}$ value.
\end{itemize}
\end{quote}

\begin{table} [ht] 
\centering
    \begin{subtable}{\textwidth}
	\centering
    \begin{filecontents*}{anscombe_r.csv}
Pearson,Spearman,bend,bicor,skipped,RIN,Bayes,dcal
0.816 (0.381-0.962),0.818,0.814,0.823,0.816,0.804,0.564,0.626
0.816 (0.019-0.994),0.691,0.804,0.805,0.759,0.645,0.564,0.589
0.816 (0.904-1),0.991,1,0.976,0.816,0.964,0.564,0.403
0.817 (NA-NA),0.5,NA,0.652,0.816,0.567,0.564,0
\end{filecontents*}
\csvautotabular[respect all]{anscombe_r.csv}
    \caption{$r$-values.}
	\end{subtable}

	\begin{subtable}{\textwidth}
	\centering
    \begin{filecontents*}{anscombe_p.csv}
Pearson,Spearman,bend,bicor,RIN,Bayes,dcal
0.002 (0.009),0.004,0.002,0.002,0.003,0.079,0.039
0.002 (0.046),0.023,0.003,0.003,0.032,0.08,0.057
0.002 (0),0,0,0,0,0.079,0.219
0.002 (NA),0.117,NA,0.03,0.069,0.079,0.5
\end{filecontents*}
\csvautotabular[respect all]{anscombe_p.csv}
    \caption{$p$-values.}
	\end{subtable}
\caption{Results of different correlation methods for the Anscombe's quartet dataset.}
\label{table_anscombe_r_p}
\end{table}

This shows that data calibration, i.e. replacing each sample with its out-of-sample prediction, reveals the underlying \emph{predictive information}. 
When calibrated using the linear regression model, the predicted values are better suited for testing the linear relationship. 
Specifically, it identifies the true relationship (Fig. \ref{fig_anscombe}, A) and also refutes the false ones (Fig. \ref{fig_anscombe}, B-C). 
This suggests that the \texttt{dcal} test can reveal the \emph{true} linear relationship between the variables and also qualify its statistical significance.

The Anscombe's dataset D exposes a critical aspect in the \texttt{dcal}-based test by providing a counter-intuitive result. 
This is because of a pitfall in the cross-validation procedure which artificially creates negative correlation when there lack of predictability. 
To the best of out knowledge, this issue has not been part of a publication but has been discussed in on-line forums: \emph{Perils of LOO corssvalidation}\footnote{http://www.russpoldrack.org/2012/12/the-perils-of-leave-one-out.html}\textsuperscript{,}\footnote{https://not2hastie.tumblr.com/post/56630997146/i-must-confess-i-was-surprised-by-the-negative}. 
The counter-intuitive result on the dataset D clarifies that this issue must be tackled while designing the \texttt{dcal} correlation test.
Here we added a simple heuristic that enforces that the original correlation sign is retained after the \texttt{dcal} test.

Based on eq. \eqref{eq1}, eq. \eqref{eq2} and the insights from the Anscombe's quartet, we devised an easy to implement algorithm for \texttt{dcal} correlation test (Algo. \ref{algo}). 

\begin{algorithm}[!htbp]
\SetAlgoLined
\SetKwInOut{Input}{input}
\SetKwInOut{Output}{output}
\Output{$r,\: p,\: r_{dcal},\: p_{dcal}$}
\Input{$x,\: y,\: \alpha,\: flag\_fast$}
 $r,\: p = cor(x,\: y)$\;
 $r_{dcal},\: p_{dcal} = 0.0,\: 0.5$\;
 \If{$\neg{flag\_fast} \; || \; p < \alpha$} {
	\For{$i=1$ \KwTo $n$}{
 		$\hat{y_i} = OLS^{(y \sim x)_{-i}}(x_i)$\;
		$\hat{x_i} = OLS^{(x \sim y)_{-i}}(y_i)$\;
   	}
	$r_{dcal},\: p_{dcal} = cor(\hat{x},\: \hat{y})$\;
 	\If{$sign(r_{dcal}) \neq sign(r)$}{
   		 $r_{dcal},\: p_{dcal} = 0.0,\: 0.5$\;
  	}
  }  
 \caption{\texttt{dcal} correlation test}
\label{algo}
\end{algorithm}

\begin{figure}[!htbp]
\centering
\includegraphics[width=0.8\linewidth,keepaspectratio]{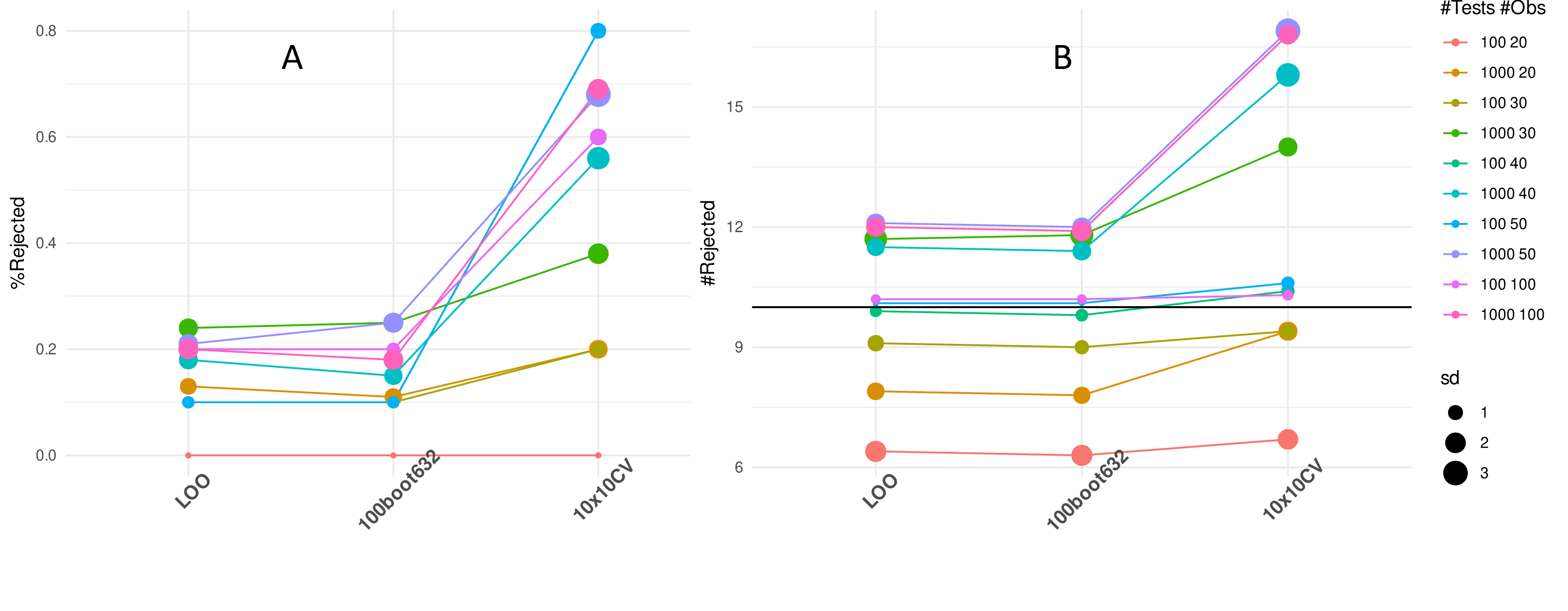}
\caption{Tests on simulated data to identify suitable out-of-sample prediction method: (A) no significant correlations expected, (B) ten significant correlations expected.}
\label{fig_oos}
\end{figure}

As there are several ways to perform OOS prediction \citep{hastie_elements_2009}, it is necessary to choose a method suitable for data calibration. 
We evaluated three OOS methods; leave-one-out (LOO) validation, 10 times repeated 10-fold cross-validation (10x10CV), and 100 times repeated bootstrap 632 (100boot632) on various settings of two types of simulated data sets, one with no relationship at all and one with some relationships. 
We found that LOO and 100boot632 performed adequately while 10x10CV lead to a higher false positive rate (Fig. \ref{fig_oos}). 
We, therefore, deemed LOO as our default choice. 
However, LOO can be computationally prohibitive with very large sample sizes and 100boot632 can serve as an alternative.

\section*{Experimental Settings and Data}
We used simulations to establish basic properties of our proposed method. For this reason we compared our method with various commonly used multiple testing correction methods which typically penalize the $p$ values based on the number of tests performed. 
These tests included, family-wise error control using the Holm's correction (\texttt{Holm}), False Discover Rate control using the Benjamini-Hochberg correction (\texttt{BH}), permutation test using the maximum statistic across the tests (\texttt{Perm\_max}) as well as permutation test for individual test (\texttt{Perm}).
We employed two $p$-value calibration methods, a first proposed by Selke and colleagues \citep{sellke_calibration_2001} (\texttt{pcalSBB}, \emph{pcal} function from the \emph{pcal} R package) and a second proposed by Bickel \citep{bickel_interval_2021} (\texttt{pcalBickel}, implemented as $pcalBickel\, =\, (1\: - abs(2.7\: p\: \ln\: p))\: p\, +\, 2\: abs(2.7\: p\: \ln\: p)$).
We also included Bayesian correlation test (the \emph{correlationBF} function from the R package \emph{BayesFactor}) which provides a Bayes factor (BF) per test and might need multiple testing correction \citep{gelman_why_2012}. 
There is no pre-defined and accepted cut-off for BF and two options have been proposed, 3 as moderate and 10 as strong evidence for accepting the $H1$  \citep{van_doorn_jasp_2020}. BF-based hypothesis testing--so called "null hypothesis Bayesian testing" (NHBT)--is prudent compared to NHST but it might be a calibration issue rather than a real disagreement between the two \citep{trafimow_2003,wetzels_2011,tendeiro_kiers_2019}. 
Instead of using the Bayes factors, we converted them to posterior probability estimates with prior probability of $0.5$ (\texttt{ppBF}, \emph{{bfactor\textunderscore to\textunderscore prob}} function from the \emph{pcal} R package) \citep{jeffreys_theory_1961,ly_harold_2016}. For all the methods statistical significance for null hypotheses rejection was set to the commonly used threshold of $\alpha=0.05$. For simulation studies we did not apply any additional correction to $p$ values from \texttt{dcal}, the two calibration methods or \texttt{ppBF}. 
This choice was made to gain a better understanding of "raw" $p$ values provides by these methods. Note that use of $\alpha=0.05$ is intensely debated (e.g., see \citep{benjamin_redefine_2018,amrhein_remove_2018,wasserstein_moving_2019}), this debate, however, is out of the scope of the current work.

The simulation studies were helpful in empirically establishing that the proposed \texttt{dcal} correlation test when applied to multiple hypotheses provides $p$ values penalized for each individual test and hence can be used without additional multiple testing correction. 
However, the simulated data can not cover all the complexities of real world data. 
We, therefore, applied the proposed test together with the comparison methods to a high-dimensional RNA-seq dataset ARXHS4 \citep{lachmann_massive_2018} to identify co-expressed genes.

\section*{Results}
\subsection*{Comparison With $p$-value Calibration and Bayesian Methods}
Using simulated data with different sample sizes and different effect sizes in the range $0.2-0.8$, we first compared our \texttt{dcal} test with two $p$-value calibration methods (\texttt{pcalSBB}, \texttt{pcalBickel}), and the posterior probability based on the Bayes factor (\texttt{ppBF}).

\begin{figure}[!htbp]
\centering
\includegraphics[width=0.8\linewidth,keepaspectratio]{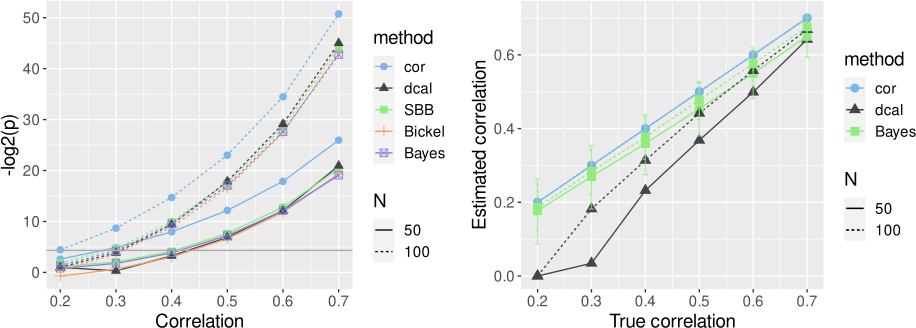}
\caption{Comparison with $p$-value calibration and Bayesian posterior probability.}
\label{fig_pcal_bayes}
\end{figure}

We observed that the \texttt{dcal} test behaved rather conservatively at lower effect sizes--i.e., high $p_{dcal}$ values--but then approaches the original $p$ values at higher effect sizes (Fig. \ref{fig_pcal_bayes}A). 
An important advantage provided by our \texttt{dcal} test is that it also provides conservative correlation value estimates at lower effect sizes where the Bayesian method remains close to the original values (Fig. \ref{fig_pcal_bayes}B). 
Similarly to the $p_{dcal}$ values the $r_{dcal}$ estimates approach the true values at higher effect sizes and with increasing sample size, a behaviour desirable in real-world applications.

\subsection*{Simulations With and Without Correlations}
\begin{figure}[ht]
\centering
\includegraphics[width=0.8\linewidth,keepaspectratio]{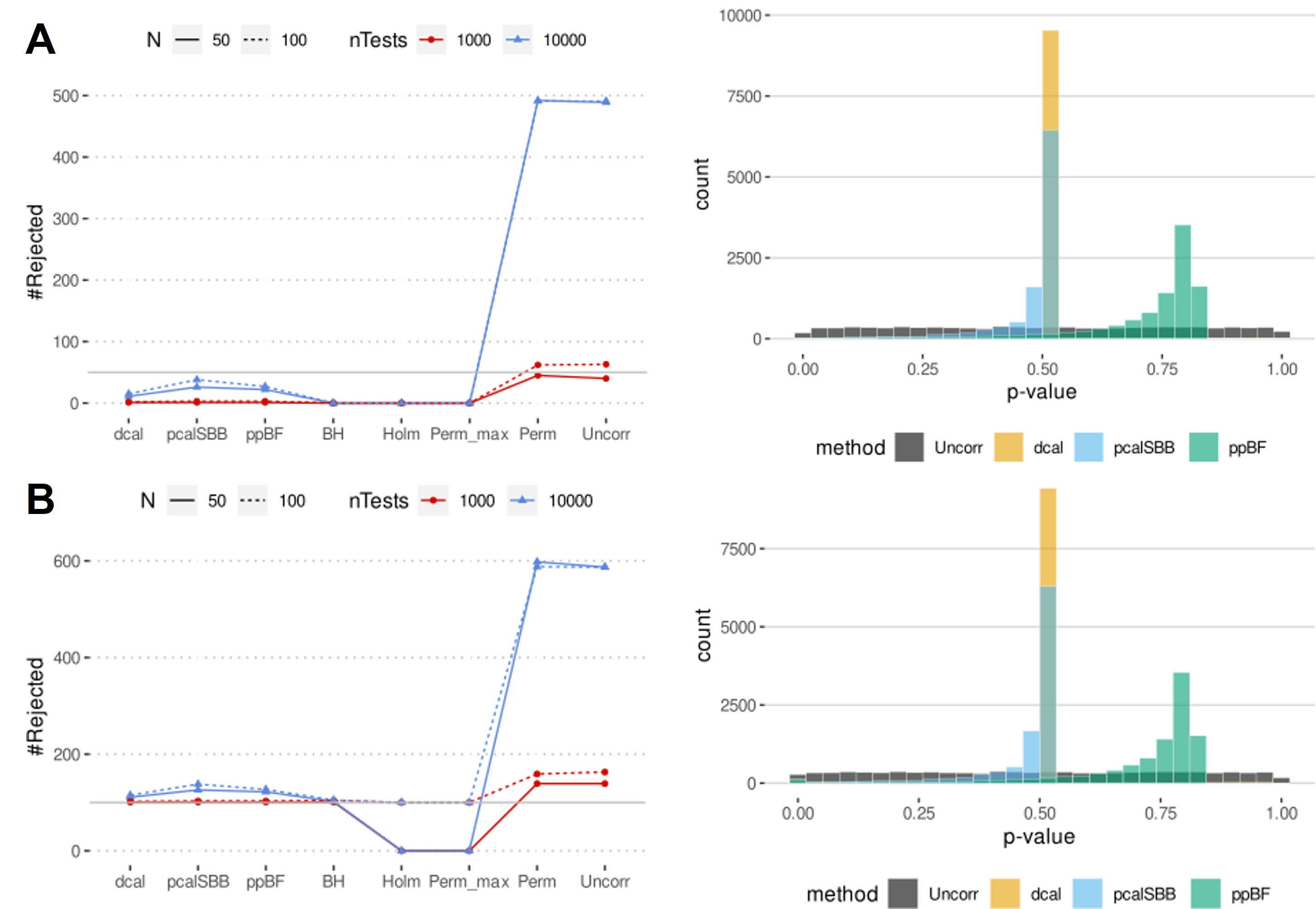}
\caption{Tests on simulated data with (A) no true correlations, and (B) 100 true correlations.}
\label{fig_sim}
\end{figure}

We sought to establish how \texttt{dcal} tests perform in a controlled scenario. 
For this we first performed a baseline simulation where there were no correlation between the variables, i.e. they were generated completely randomly using a Gaussian distribution with $\mu=0$ and $\sigma^2=1$. 
We generated 100 independent $x$ variables to test against a single dependent variable $y$, a usual setting in real-world scenarios. 
Here a \textit{well-calibrated} method should not provide any significant results. 
Indeed, \texttt{dcal} performed well and did not produce many false positives (Fig. \ref{fig_sim} A).
The same was true for other methods employing explicit multiple testing correction.

We then generated data with 100 variables correlated with the outcome variable with effect size of $0.5$. 
A good method should provide 100 rejections if it can correctly handle the multiple tests. 
We indeed observed that \texttt{dcal}, \texttt{pcalSBB}, and \texttt{FDR} generated close to 100 rejections when performing 1000 or 10000 tests (Fig. \ref{fig_sim} B). 
Two methods, \texttt{Holm} and \texttt{Perm\_max}, were overly conservative with the smaller sample size (N=50).

Overall, \texttt{Perm} and \texttt{Uncorr} produced many more false positives, and more tests (10000) resulted in more false positives, as expected.

\subsection*{Gene Co-expression Analysis}
We analysed RNA-seq data to to identify which of the available 32831 genes are significantly correlated with the human specific \textit{ARHGAP11B} that is involved neocortex expansion \citep{florin_ARHGAP11B_2015}. 

\begin{figure}[!ht]
\centering
\includegraphics[width=0.5\linewidth,keepaspectratio]{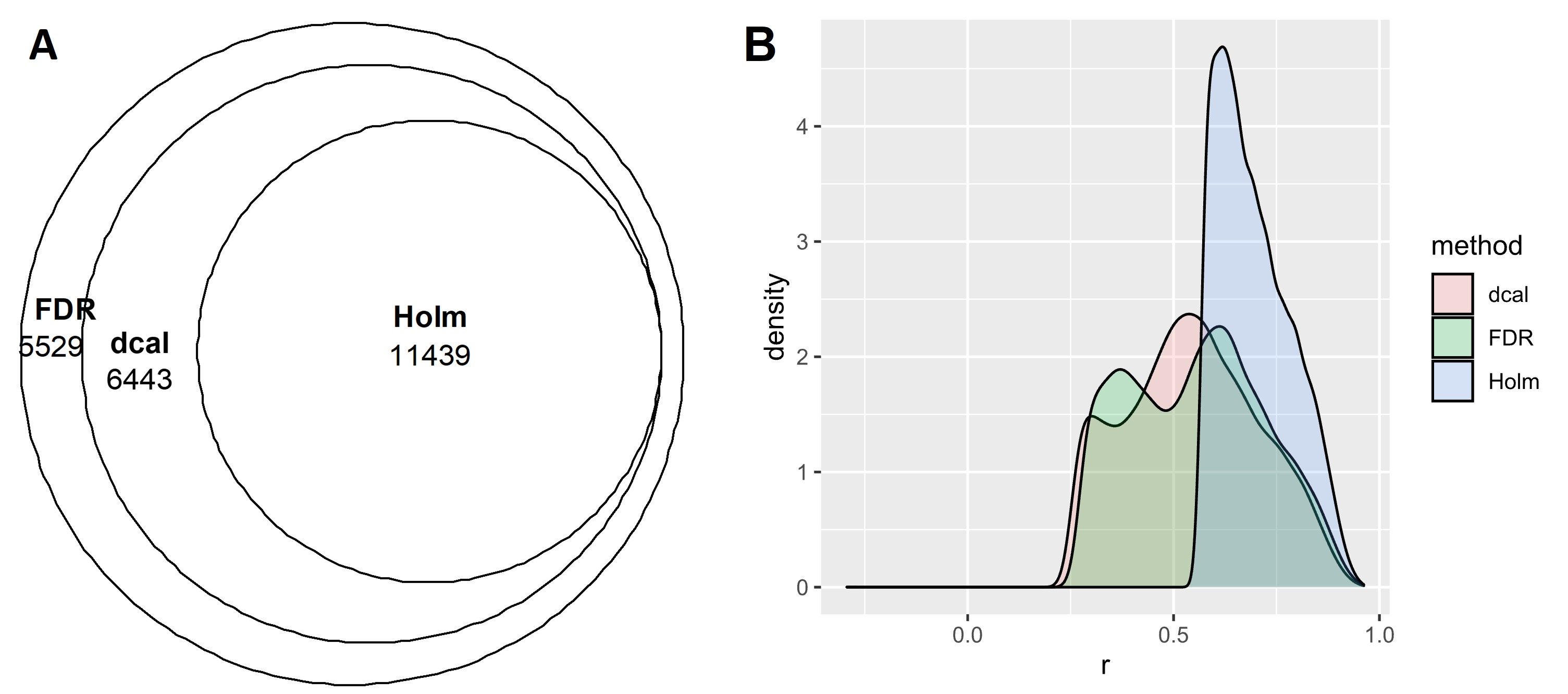}
\caption{Gene co-expression analysis of the \textit{ARHGAP11B} gene. 
(A) shows that the \texttt{dcal} method lies in-between \texttt{FDR} and \texttt{Holm} correction.
(B) shows the corresponding density plots of the $r$ values deemed significant ($\alpha = 0.05$).}
\label{fig_gene_coexpr}
\end{figure}

The three methods were consistent such that the most conservative correction \texttt{Holm} (11439 significant) was subsumed by \texttt{dcal} (17882 significant) which in turn was a subset of \texttt{FDR} (23411 significant) (\ref{fig_gene_coexpr} A).

\subsection*{Effect of outliers}
We then investigated how the \texttt{dcal} method deals with outliers in general. 
To this end, we performed simulations following previous work \citep{pernet_robust_2013,bishara_reducing_2015} and compared with a robust correlation method, \texttt{skipped} correlation \citep{wilcox_skipped_2015}.
In all cases $10\%$ of the data was sampled from the corresponding outlier distribution.

\begin{figure}[!htbp]
\centering
\includegraphics[width=0.8\linewidth,keepaspectratio]{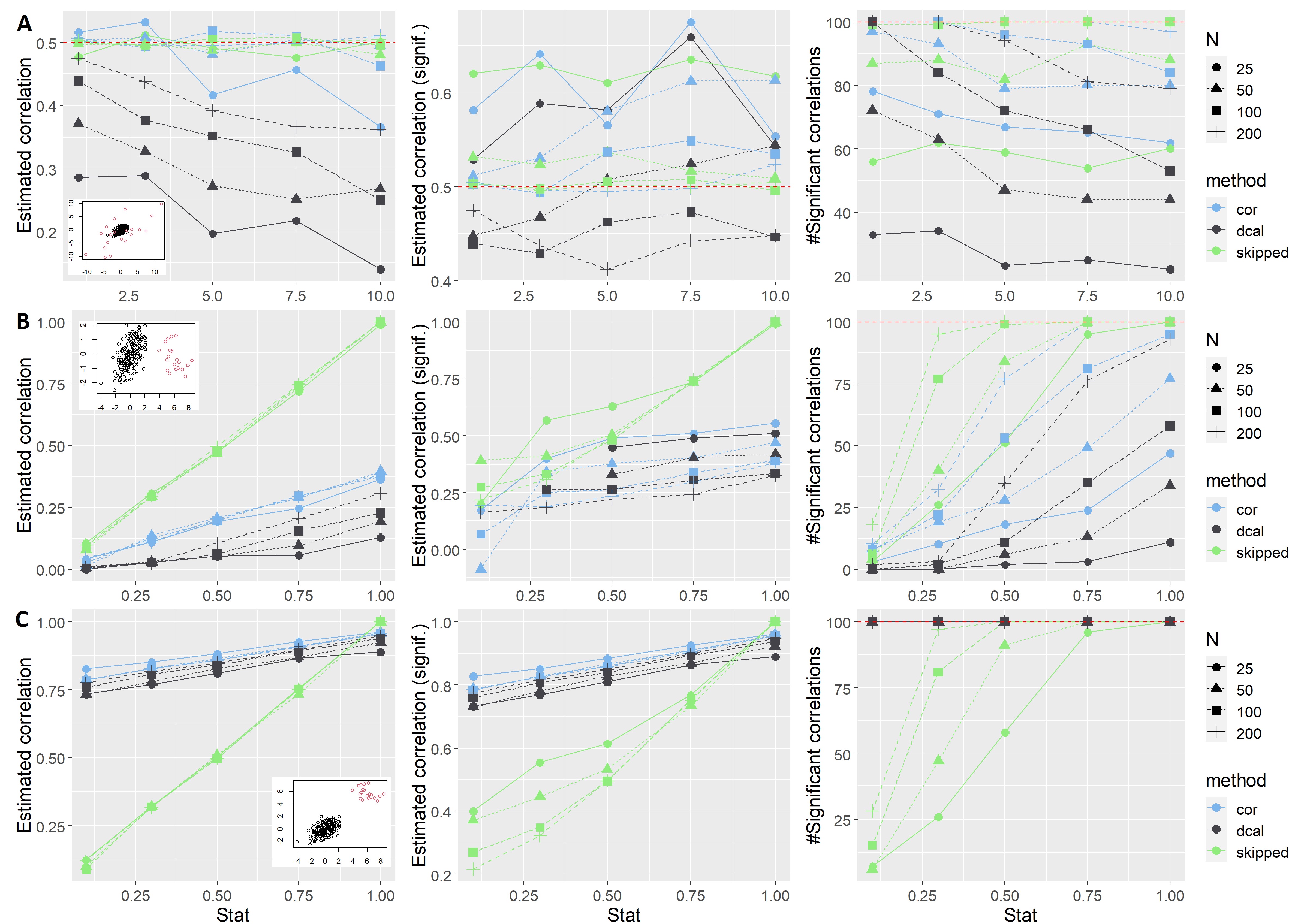}
\caption{Effect of outliers on correlation estimation. 
The three rows show different types of outliers as shown by the example insets: high variance (A), univariate (B), and bivariate (C).
The columns show different evaluations of the three kinds of outlier scenarios.}
\label{fig_sim_outlier}
\end{figure}

For the high variance outliers we varied the standard deviation of the population distributions while keeping correlation constant at $0.5$. 
With higher standard deviation of the outlier distribution \texttt{dcal} estimated correlation values dropped while $cor$ also shows a drop but only for small sample size and \texttt{skipped} gives more consistent estimates (Fig. \ref{fig_sim_outlier} A, left). 
If we look only at the significant correlations ($\alpha = 0.05$, middle panels) then we observe more fluctuations around the expected value of $0.5$ with smaller samples overestimating the $r$. 
The sensitivity of the methods (right panel) as expected was generally lower for smaller samples with \texttt{dcal} showing lowest sensitivity and \texttt{skipped} the highest. 
 
For the next two kinds of outliers (univariate and bivariate), we varied the correlation between the two variables.  
For the univariate outliers (Fig. \ref{fig_sim_outlier} B), the \texttt{skipped} correlation estimated the $r$ values close to the expected values at all sample sizes while both $cor$ and \texttt{dcal} underestimated them with the latter being more pessimistic.
Similar observations could be made while looking only at the significant correlations.
While the sensitivity of all the methods increases with increasing value of true $r$, \texttt{skipped} was most sensitive followed by $cor$ and \texttt{dcal}.

With bivariate outliers (Fig. \ref{fig_sim_outlier} C), some of the aforementioned trends reversed.
\texttt{skipped} estimates of $r$ were close to the true value while both $cor$ and \texttt{dcal} overestimated them.
The sensitivity of \texttt{skipped} was lower while the other two methods showed near-perfect sensitivity.

In line with the Anscombe's quartet analysis (Fig. \ref{fig_anscombe}), \texttt{dcal} showed reduced sensitivity in presence of outliers with reduced effect sizes. 
However, with bivariate outliers \texttt{dcal} behaved similar to $cor$ suggesting that the type of outliers can impact its behaviour, which can be expected of any method.
The reduced sensitivity can be a desirable property to avoid false positives in presence of outliers. 

\section*{Discussion and Conclusions}
We proposed a novel ML-based predictive method to perform linear correlation NHST. 
Our method, \texttt{dcal}, calibrates the data using the alternative hypothesis model (here linear regression) such that each calibrated data point encodes information regarding its conformation to that hypothesis. 
Subsequent application of the conventional test to this calibrated data yields calibrated test results. 
Importantly, both the correlation estimate ${r_{dcal}}$ and the $p$-value ${p_{dcal}}$ are calibrated. 
This provides an advantage compared to the probability calibration methods which only calibrate the $p$ values. 

Analysis of the Anscombe's quartet dataset show that \texttt{dcal}-based correlation test can handle scenarios where both probability calibration and Bayesian methods fail (Fig. \ref{fig_anscombe}, Table \ref{table_anscombe_r_p}).
These results are further validated by several simulations as well as real-world gene co-expression data analysis.
Our results regarding multiple testing suggest that the output of \texttt{dcal} does not require additional correction for multiple tests. 

Our method has important implications for scientific and industrial applications. 
A major limitation of traditional multiple testing correction methods is that they become conservative with increasing number of tests. 
As the number of variables keep growing, e.g. number of probes in a micro-array experiment, resolution of neuorimgaing and satellite data, traditional frameworks become too conservative.
\texttt{dcal} can potentially alleviate this limitation as ${p_{dcal}}$ can be used directly for qualifying a test as significant or not.
Taken together, our proposed method can help scientific investigation by providing an easy to use alternative to traditional linear correlation analysis and multiple testing correction.

Another important consideration in the big data era is the dependence of $p$ values on the sample size \citep{gomez_2021}. As increasing sample size leads to decreasing $p$-value it also increases the chances of false positives. How this dependence plays out with \texttt{dcal} needs further investigation.

Statistics is a founding discipline of machine learning and many algorithms and theoretical results have benefited from a close interaction with statistics. 
In this work, we use ML to make a contribution to statistics. 

\subsection*{Limitations and Future Work}
Here we developed a data-calibration-based test for linear correlation. 
However, the proposed \texttt{dcal} paradigm provides a framework which can be extended to other statistical tests using appropriate algorithms, for instance t-test using logistic regression. 
We envision a general data-calibration-based NHST framework applicable to various statistical tests which we fondly call Significance testing INcorporating Generalization (\texttt{SING}).

Our current algorithm (Algo. \ref{algo}) uses LOO validation for obtaining out-of-sample predictions. 
It is possible to use other schemes like bootstrap to obtain out-of-sample predictions but here we used LOO as it provides the most unbiased estimates which is beneficial for our purpose. 
However, LOO can be computationally expensive for large data sets and further tests should be performed to identify optimal choices.

The $p_{dcal}$ values were observed to be close to that of $p$ value calibration methods. 
Given the low computational requirement of $p$ value calibration, they offer a tempting preference. 
However, they do not perform well on corner-cases---as we showed using the Anscombe's quarter---and they do not offer calibrated estimates of the statistic estimate. 
Investigating similarities and differences between these two approaches may provide novel insights and possibly faster algorithms.

We acknowledge that our use of Bayesian methods is not Bayesian in spirit as we converted the Bayes factors into posterior probability estimates and that too with a fixed prior probability of $0.5$. 
Although resulting point estimates are non-Bayesian in spirit they serve well the comparative purpose as we intended in this study. 

Taken together, our proposed method provides an alternative way to look at statistical significance testing. We hope that the conceptual and theoretical novelty helps in developing new methods making the application of the widely used NHST framework more robust and less prone to misinterpretation.

\section*{Acknowledgements}
We cordially thank Prof. Carlos Soares (U. Porto) for fruitful brainstorming sessions. We also thank Prof. Maya Mathur (Stanford), and Prof. Wolfgang Huber (EMBL) for their comments on the initial ideas of this research.

This work was partly supported by the Deutsche Forschungsgemeinschaft (DFG, PA 3634/1-1 and EI 816/21-1), and the Helmholtz Portfolio Theme “Supercomputing and Modelling for the Human Brain”.

\bibliographystyle{unsrtnat}
\bibliography{sing-bib}

\end{document}